\documentclass[aps,prl,superscriptaddress,twocolumn,showpacs]{revtex4}

\renewcommand*{\[}{\begin{equation}}

\renewcommand*{\]}{\end{equation}}

\newcommand{\myscalebox}[1]{\scalebox{0.8}[0.8]{#1}}
\newcommand{\myscaleboxa}[1]{\scalebox{0.55}[0.55]{#1}}
\newcommand{\myscaleboxb}[1]{\scalebox{0.35}[0.35]{#1}}

\usepackage{epsfig}

\usepackage{hyperref}

\begin{document}

\title{Separation of Target Structure and Medium Propagation Effects in High-Harmonic Generation}

\author{Cheng Jin}

\affiliation{J. R. Macdonald Laboratory, Physics Department, Kansas
State University, Manhattan, Kansas 66506-2604, USA}

\author{Hans Jakob W\"{o}rner}

\affiliation{Joint Laboratory for Attosecond Science, National
Research Council of Canada and University of Ottawa, 100 Sussex
Drive, Ottawa, Ontario, Canada K1A 0R6}

\author{V. Tosa}

\affiliation{National Institute for R$\&$D of Isotopic and Molecular
Technologies, 400293 Cluj-Napoca, Romania}

\author{Anh-Thu Le}
\affiliation{J. R. Macdonald Laboratory, Physics Department, Kansas
State University, Manhattan, Kansas 66506-2604, USA}

\author{Julien B. Bertrand}
\affiliation{Joint Laboratory for Attosecond Science, National
Research Council of Canada and University of Ottawa, 100 Sussex
Drive, Ottawa, Ontario, Canada K1A 0R6}

\author{R. R. Lucchese}
\affiliation{Department of Chemistry, Texas A$\&$M
University, College Station, Texas 77843-3255, USA}

\author{P. B. Corkum}
\affiliation{Joint Laboratory for Attosecond Science, National
Research Council of Canada and University of Ottawa, 100 Sussex
Drive, Ottawa, Ontario, Canada K1A 0R6}

\author{D. M. Villeneuve}
\affiliation{Joint Laboratory for Attosecond Science, National
Research Council of Canada and University of Ottawa, 100 Sussex
Drive, Ottawa, Ontario, Canada K1A 0R6}

\author{C. D. Lin}
\affiliation{J. R. Macdonald Laboratory, Physics Department, Kansas
State University, Manhattan, Kansas 66506-2604, USA}

\date{\today}

\begin{abstract}
We calculate high-harmonic generation (HHG) by intense infrared
lasers in atoms and molecules with the inclusion of macroscopic
propagation of the harmonics in the gas medium. We show that the
observed experimental spectra can be accurately reproduced
theoretically despite that HHG spectra are sensitive to the
experimental conditions. We further demonstrate that the simulated
(or experimental) HHG spectra can be factored out as a product of a
``macroscopic wave packet" and the photo-recombination transition
dipole moment where the former depends on the laser properties and
the experimental conditions, while the latter is the property of the
target only. The factorization makes it possible to extract target
structure from experimental HHG spectra, and  for ultrafast dynamic
imaging of transient molecules.
\end{abstract}


\maketitle

High-harmonic generation (HHG) has been employed to probe electronic
structure of molecules on an ultrafast time scale in recent years
\cite{itatani-nature-2004,smirnova-nature-2009,hans-nature-2010}.
When molecules are placed in an intense laser field, electrons that
are removed earlier may be driven back to recollide with the parent
ion. HHG occurs when the returning electrons recombine with the
parent ion with the emission of high-energy photons as in an inverse
photoionization (PI) process. Since PI is a sensitive tool for
probing electronic structure of molecules, HHG may serve likewise,
but with the advantage of ultrafast temporal resolution as well as
covering a coherent broad spectral range from XUV to soft-X-rays.
Experimentally, however, HHG is generated from all the molecules in
the interaction region. The radiations from them co-propagate with
the fundamental infrared (IR) beam nonlinearly. To extract structure
information of individual molecules, e.g., the amplitude and phase
of PI transition dipole from the measured HHG, the propagation
effect in the medium should be investigated. For molecular targets,
this has not been done so far. Instead, it was often assumed that
HHG was measured under the perfect phase-matching conditions and
that the observed harmonics were directly proportional to the
harmonics from a single molecule. While such assumptions may be
adequate for explaining many experimental observations
qualitatively, such as the dependence of HHG on molecular alignment
and on symmetry of the molecular orbital, the two-center
interference \cite{Kanai-nature-05,Lein-prl-02}, and
multiple-orbital contributions to HHG \cite{McFarland-sci-08}, they
are inadequate if accurate structure information of individual
molecules are to be extracted from the observed HHG spectra.

The effect of macroscopic propagation on the observed HHG spectra
for atoms has been investigated extensively in the past two decades
in connection with the generation of attosecond pulses. However, we
are not aware of any direct comparison between experimental HHG
spectra and theoretical simulations over an extended spectral
region. The Maxwell's equations that govern the propagation of the
fundamental driving IR field and the generated harmonics are well
established. To carry out such propagation calculations, accurate
induced atomic dipoles generated by lasers for hundreds of peak
intensities should be calculated which serve as the source term of
the harmonics. These induced dipoles are often calculated using the
strong-field approximation (SFA), or the so-called Lewenstein model
\cite{lew-pra-94}. The SFA does not describe the laser-atom
interactions accurately, thus the results from the propagation can
only be used to qualitatively interpret the experiments. While
accurate induced dipoles can be obtained from solving the
time-dependent Schr\"{o}dinger equation (TDSE), the calculation is
rather time consuming and was rarely attempted except for very few
occasions \cite{mette-pra-06}. Thus after two decades, our
understanding of experimental HHG data is still mostly at the
qualitative level.

\begin{figure}
\mbox{\rotatebox{0}{\myscalebox{
\includegraphics[trim=65mm 65mm 45mm 15mm, clip]{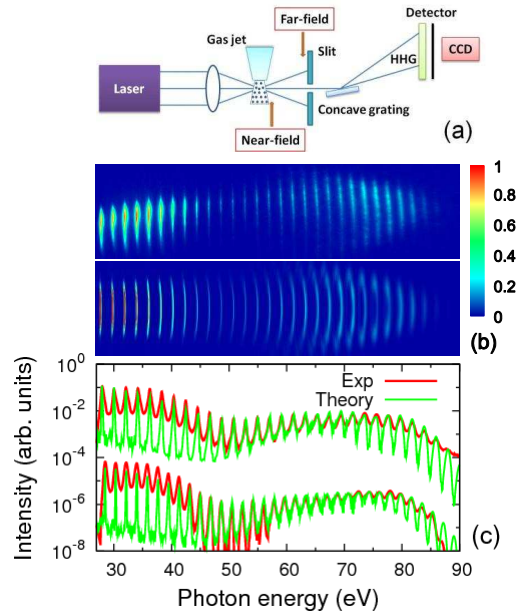}}}}
\caption{(Color online)  (a) Typical configuration for measuring HHG
in the far field. (b) HHG spectra of Ar generated by 1200 nm laser.
Upper frame: experiment; lower frame: theory. (c) Comparison of
theoretical (green curves) and experimental (red curves) HHG yields
integrated over the vertical dimension for 1200 nm (upper curves)
and 1360 nm (lower curves) lasers. Laser parameters are given in the
text.\label{Fig1}}
\end{figure}

In this Letter, we show that such limitations have been removed. We
generate HHG spectra theoretically under experimental conditions and
the results are compared directly to the observed data. The
simulated spectra agree well with the measured one, over a broad
photon energy region. The experiments were taken in a geometry as
depicted in Fig.~\ref{Fig1}(a), using IR laser pulses with
wavelengths of 1200 nm and 1360 nm, respectively, generated in a
high-energy optical parametric amplifier (HE-TOPAS). The TOPAS was
pumped by the output of a multipass femtosecond amplifier system (8
mJ, 32 fs, 50Hz). High-harmonic radiation generated from the gas jet
was allowed to propagate and a slit was placed downstream. After the
slit, a concave grating dispersed the harmonics which were then
detected with a CCD detector. The gas jet was formed from a
supersonic expansion of Ar or N$_2$ at a stagnation pressure of 3
bars. Experiments were performed on either isotropic or aligned
N$_2$ molecules. The experimental spectra were corrected for the
response of the grating and detector.

Fig.~\ref{Fig1}(b) shows the HHG spectra of Ar generated by a 1200
nm laser. The horizontal axis is the photon energy and the vertical
axis is the transverse spatial dimension. The upper frame is from
the measurement, while the bottom frame is from the simulation. The
two spectra are normalized to each other at harmonic 75, or at
photon energy of 77 eV. There is a general agreement between the two
spectra. The ``up-down" asymmetry in the experimental HHG spectra is
due to asymmetry in the laser beam profile. The faint features near
50 eV are the ``famous" Cooper minimum in Ar \cite{Cooper-pra-62},
observed in photoionization, as well as in earlier HHG spectra
\cite{hans-prl-2009,Minemoto-pra-08}. The harmonic yields integrated
over the vertical dimension are compared in the upper half of
Fig.~\ref{Fig1}(c). The lower half shows the HHG spectra taken with
the 1360 nm laser. In both cases, we can see very good agreement
between theory and experiment over the 30-90 eV region covered.
Experimentally, the gas jet is 0.5 mm long and placed 3 mm after the
focus. A vertical slit with a diameter of 100 $\mu$m is placed 24 cm
after the gas jet. For the 1200 (1360) nm the beam waist at the
laser focus is 47.5 (52.5) $\mu$m, and the pulse duration is
$\sim$40 ($\sim$50) fs. To achieve best overall agreement, in the
simulation the peak intensity and gas pressure for each wavelength
are adjusted till best overall fit in the data are achieved. Thus
for 1200 nm laser, the peak intensity for experiment (theory) is 1.6
(1.5)$\times$10$^{14}$W/cm$^{2}$, gas pressure is 28 (84) Torr. For
the 1360 nm laser, the corresponding parameters are 1.25
(1.15)$\times$10$^{14}$W/cm$^{2}$, and 28 (56) Torr, respectively.
By using a higher pressure in the simulation, we find that the
higher harmonics become sharper, as in experiments. The (normalized)
envelope of the harmonic spectra, however, does not depend much on
the gas pressure, see Fig.~\ref{Fig3}(b) below.

In the theoretical simulation, we first obtain single-atom induced
dipole using the quantitative rescattering (QRS) theory
\cite{toru-2008,at-pra-2009,CDL-10}. The Ar is treated in the
single-active electron approximation using model potential proposed
by Muller \cite{Muller}. The resulting induced dipoles for different
peak intensities are then fed into the Maxwell's equations. The
propagation equations for the fundamental field and the harmonics
are the standard ones
\cite{Priori-pra-2000,Mette-jpb,tosa-pra-2005}. For Ar target, we
include dispersion, absorption, Kerr and plasma effects on the
fundamental field in the medium. For the harmonics, only the
dispersion and absorption are included. The harmonic yields emitted
at the exit face of the gas jet (near field) are propagated to the
far field where the harmonics are measured. They are obtained from
the near-field harmonics through a Hankel transform
\cite{L'Huillier-1992,tosa-2009}. We assume the laser beam in the
entrance of gas jet has the Gaussian shape.

A careful examination of Figs.~\ref{Fig1}(b) and (c) reveals that
there are still small discrepancies between the experimental data
and the simulation. The harmonic width (or harmonic chirp) is
narrower from the theory than from the experiment. Harmonic chirp is
a direct consequence of temporal variation of laser intensity. The
harmonic width is mainly influenced by the pulse duration, pressure,
and laser intensity \cite{Mette-pra-99,Zair-prl-08,He-pra-09}. The
width decreases with increasing pulse duration, and with decreasing
gas pressure. Other experimental factors like use of the slit and
location of the detector also can affect the HHG spectra.

\begin{figure}
\mbox{\rotatebox{270}{\myscaleboxa{
\includegraphics{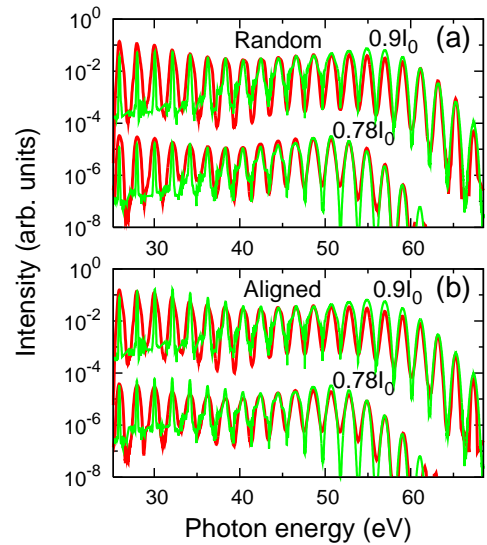}}}}
\caption{(Color online)  Comparison of HHG spectra from theory
(green curves) and experimental results (red curves) of Ref.
\cite{hans-prl-2010}, (a) for randomly distributed N$_2$ and (b) for
N$_2$ aligned along laser polarization direction. The laser
intensities are indicated where $I_0$=10$^{14}$ W/cm$^2$. See text
for additional laser parameters.\label{Fig2}}
\end{figure}

\begin{figure}
\mbox{\rotatebox{270}{\myscaleboxb{
\includegraphics{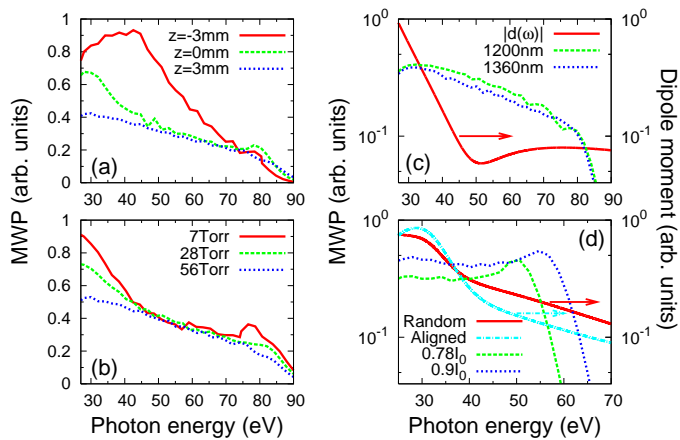}}}}
\caption{(Color online)  (a) Dependence of macroscopic wave packet
$|W(\omega)|$ (MWP) on the position of the Ar gas jet with respect
to laser focus; (b) Effect of gas pressure on MWP. The curves are
renormalized such that they should fall on the same curve if the
perfect phase-matching condition is fulfilled. (c) The MWP for the
two lasers, and the magnitude of the PI transition dipole moment of
Ar. (d) Same as in (c) but for N$_2$. The MWP is for two different
laser intensities, and the averaged PI transition dipoles are for
isotropic and aligned molecules. The laser intensities are indicated
where $I_0$=10$^{14}$ W/cm$^2$. See text.\label{Fig3}}
\end{figure}

High-order harmonics from molecules by 1200 nm lasers have been
reported for aligned and randomly distributed N$_2$ and CO$_2$
recently \cite{hans-prl-2010}. Here we report our simulated results
for N$_2$, at the two peak laser intensities, 0.9 and
1.1$\times$10$^{14}$W/cm$^{2}$, reported in \cite{hans-prl-2010}. To
achieve good agreement in the cutoff positions, the two intensities
used in the theory are 0.78 and 0.9$\times$10$^{14}$W/cm$^{2}$
instead, respectively. Since the experiment was carried out at low
laser intensity and low gas pressure, the harmonics are propagated
without absorption and dispersion effects from the medium, and the
fundamental laser field is not modified through the medium
\cite{jin-2009}. In the theoretical simulation, we first obtain
induced dipoles of fixed-in-space molecules using QRS theory
\cite{at-pra-2009,CDL-10} for different laser peak intensities. The
induced dipoles are averaged coherently according to the alignment
distribution and then fed into the Maxwell's equations.
Fig.~\ref{Fig2} shows the good overall agreement between the
measured and the simulated spectra, for both randomly distributed
and aligned N$_2$. By examining the experimental HHG spectra more
carefully, they reveal a shallow minimum at 38$\pm$2eV (low
intensity) and at 41$\pm$2eV (high intensity) for both aligned and
unaligned molecules. The theory also predicts a minimum: for
unaligned molecules, the minimum is at $\sim$39eV for low intensity
and $\sim$40eV for high intensity. For aligned molecules, the
minimum is at $\sim$42eV for low intensity and $\sim$44eV for high
intensity. In the experiment, the degree of alignment was estimated
to be $\langle \cos^{2}\theta \rangle$=0.6-0.65. In the simulation,
an alignment distribution of $\cos^{4}\theta$ is used. Note that
only HOMO is included in the calculation. We believe that this is
the first time that HHG spectra from molecules have been calculated
including the propagation effect in the medium and the simulated
results have been compared directly to the measured spectra. In the
future, HHG spectra taken at different alignment angles should be
compared together. Such comparison would help to identify factors
that contribute to the remaining discrepancies between experiment
and simulation.

The macroscopic HHG spectra can be expressed as \cite{jin-2009}
\begin{eqnarray}
\label{mwp}S_{h}(\omega)\propto\omega^{4}|W(\omega)|^{2}|d(\omega)|^{2}
\end{eqnarray}
where $W(\omega)$ (the complex amplitude) is called ``Macroscopic
wave packet" (MWP), $d(\omega)$ is PI transition dipole moment for
the atom. For the molecule $d(\omega)$ is taken to be the coherently
averaged PI transition dipole moment
$d^{avg}(\omega)=\int_{0}^{\pi}N(\theta)^\frac{1}{2}\rho(\theta)d(\theta,\omega)\sin\theta
d\theta$, where $N(\theta)$ is the alignment-dependent ionization
probability, $\rho(\theta)$ is the alignment distribution, and
$d(\theta,\omega)$ is the parallel component of the
alignment-dependent transition dipole moment \cite{at-pra-10}. The
polarization of pump laser is assumed parallel to probe laser. For
unaligned molecules, $\rho(\theta)$ is a constant. Actually, MWP has
the clear physical meaning. It can be considered as the collective
effect of microscopic wave packets for the returning electrons
\cite{at-pra-2009,CDL-10}, which is governed by Maxwell's equations.
In other words, the laser and macroscopic medium effects are all
combined into MWP.

The validity of Eq.~(\ref{mwp}) has been checked in Jin {\it et al.}
\cite{jin-2009} when both the laser intensity and the gas pressure
are low. The correctness of this relation has been assumed in
Itatani {\it et al.} \cite{itatani-nature-2004} by comparing Ar with
N$_2$, and in Levesque {\it et al.} \cite{Lesvque-prl-2007} for rare
gas atoms. However, this relation has not been carefully checked for
different focusing conditions and laser parameters. Theoretically,
we have checked the validity of Eq.~(\ref{mwp}) carefully. We
carried out macroscopic propagation calculation of HHG using
single-atom (single-molecule) induced dipole obtained by QRS and by
SFA. We have been able to show that the MWP obtained from the two
calculations agree rather well, irrespective of laser parameters or
the focusing conditions. In other words, the medium propagation only
affects HHG through its modifications on the MWP. In this way, to
study propagation effect on HHG, we can just study how the MWP (only
the amplitude is considered below) depends on the lasers and the
experimental conditions.

In Fig.~\ref{Fig3}(a) we show the dependence of MWP on the position
of the Ar gas jet with respect to the laser focus. The laser
intensity is 1.6$\times$10$^{14}$W/cm$^{2}$, and the gas pressure is
56 Torr. For easy visualization we show the smooth envelope of
$|W(\omega)|$. The three curves are for the gas jet at z=-3 mm
(laser focus before gas jet), z=0 (at) , +3 (after). It is generally
known that HHG achieves best phase-matching if the gas jet is placed
behind the laser focus where the dipole phase from the harmonic can
be partially canceled by the Gouy phase. Thus among the three
curves, the ``after" curve is the flattest one. For the ``before"
focus, the MWP varies most as the photon energy is changed, and the
phase (not shown) varies widely from order to order --- reflecting
poor phase-matching for this geometry.

In Fig.~\ref{Fig3}(b) we compare the MWP derived from changing the
Ar gas pressure for the ``after" focusing condition. The MWP has
been normalized by the ratio of the pressure. Under perfect
phase-matching condition, the MWP $|W(\omega)|$ (the amplitude)
should be proportional to the pressure
\cite{VS-OE-07,Shiner-prl-09}. The three curves are on top of each
other from 45-75 eV, indicating good phase-matching in this energy
region,  but differ somewhat at lower and higher energies,
indicating good phase-matching condition is not fulfilled. This
demonstrates that phase-matching condition cannot be achieved for
all the harmonics in a given experiment.

According to Eq.~(\ref{mwp}), the minimum in the HHG spectra can
occur for different reasons. In Fig.~\ref{Fig3}(c), the MWP derived
from Ar target using 1200 nm and 1360 nm lasers are shown. The two
MWP's are quite similar but near 50 eV, they have slight different
slopes. On the other hand, the PI transition dipole reveals a clear
but broad Cooper minimum near 50 eV. Thus the broad minimum in the
HHG spectra shown in Fig.~\ref{Fig1}(c) is due to the minimum in the
PI transition dipole. To pin down the position of the ``real"
minimum, on the other hand, is not as easy since the minimum
position can be modified somewhat by the MWP.

Similar analysis can be carried out on the HHG spectra of N$_2$
shown in Fig.~\ref{Fig2}. The averaged PI transition dipole indeed
shows a rapid drop near 40 eV, which is due to the presence of a
shape resonance \cite{Lucchese-pra-82} of N$_2$ in the lower energy.
The rapid drop is more pronounced for aligned molecules than for
random ones, see Fig.~\ref{Fig3}(d). For the MWP, under the same
laser intensity, we have checked that they are the same for randomly
distributed and aligned molecules. Thus it explains why the HHG from
single-molecule response can be used to interpret how the intensity
of each harmonic changes with pump-probe time delay in Le {\it et
al.} \cite{at-prl-09}. However, the MWP changes more rapidly with
laser intensity, especially for the longer wavelength laser used
here. We note that the two MWP's in Fig.~\ref{Fig3}(d) have somewhat
different slopes near 40 eV. The multiplication of the MWP and the
PI transition dipole results in a weak minimum in the observed HHG
spectra. The minimum would be more clearly seen if the molecules
were better aligned.  From  Le {\it et al.} \cite{at-pra-2009}, the
minimum in PI transition dipole changes rapidly with the alignment
angle and the effect is severely averaged out when molecules are not
well aligned. We further mention that the MWP in Figs.~\ref{Fig3}(c)
and (d) are rather different. They are due to the large difference
in the laser peak intensities used \cite{jin-2009}. In the future,
it is desirable that predictions such as those in Fig.~\ref{Fig3} be
checked experimentally.

In summary, we demonstrated that experimental HHG spectra can now be
accurately reproduced from {\it ab initio} calculations. The theory
starts with the calculation of laser-induced dipole from single atom
or molecule using the recently developed quantitative rescattering
(QRS) theory \cite{at-pra-2009,CDL-10}. The propagation effect of
the fundamental field and the harmonics in the medium is
incorporated by solving the Maxwell's equations. We further showed
that the simulated (and experimental) HHG spectra can be expressed
as the product of a ``macroscopic wave packet" (MWP) and the
photo-recombination transition dipole moment. The latter is a
property of the target, and is independent of the lasers, nor of the
propagation effect. This factorization makes it possible to extract
target structure information from the experimental HHG spectra. It
provides the needed theoretical basis for using HHG as ultrafast
probes of excited molecules, such as those demonstrated recently
\cite{hans-nature-2010}. Clearly, this work also opens up
opportunities for the quantitative studies of the phases of HHG
which are fundamental to the generation of attosecond pulses.

This work was supported in part by Chemical Sciences, Geosciences
and Biosciences Division, Office of Basic Energy Sciences, Office of
Science, U.S. Department of Energy.

\end{document}